\documentclass[aps,prc,reprint,superscriptaddress,nofootinbib]{revtex4-2}
\usepackage{graphicx}
\usepackage{dcolumn}
\usepackage{bm}
\usepackage{epstopdf}
\usepackage{amsmath}
\usepackage{hyperref}
\usepackage{color}
\usepackage{cancel}
\usepackage{booktabs}

\begin{document}

\title{Tilted-axis-cranking covariant density functional theory for the high-spin spectroscopy in $^{69}$Ga}

\author{Y. P. Wang}
\affiliation{State Key Laboratory of Nuclear Physics and Technology, School of Physics, Peking University, Beijing 100871, China}
\author{Y. K. Wang}
\affiliation{State Key Laboratory of Nuclear Physics and Technology, School of Physics, Peking University, Beijing 100871, China}
\author{P. W. Zhao}
\email{pwzhao@pku.edu.cn}
\affiliation{State Key Laboratory of Nuclear Physics and Technology, School of Physics, Peking University, Beijing 100871, China}

\date{\today}

\begin{abstract} 

The tilted-axis-cranking covariant density functional theory is applied to investigate the three newly-observed positive-parity bands S\uppercase\expandafter{\romannumeral1}, S\uppercase\expandafter{\romannumeral2}, and S\uppercase\expandafter{\romannumeral3} in $^{69}$Ga.
The energy spectra and angular momenta are calculated and compared with the experimental data.
For the yrast band S\uppercase\expandafter{\romannumeral1}, pairing correlations play a crucial role for the states with spin $I\leq 23/2\hbar$. 
The bands S\uppercase\expandafter{\romannumeral2} and S\uppercase\expandafter{\romannumeral3} are suggested to be signature partner bands with positive and negative signatures, respectively.
The transition probabilities $B(E2)$ for these bands are predicted, and await further experimental verification.
By analyzing the angular momentum alignments microscopicly, it is revealed that the $g_{9/2}$ protons and neutrons play an important role in the description of the collective structure of $^{69}$Ga.

\end{abstract}

\maketitle

\date{today}

\section{Introduction}

The spectroscopic properties of nuclei in the $A\approx 60-70$ mass region have caught much attention of experimental and theoretical studies in the past few decades.
They provide an ideal laboratory for the study of the interplay between the single-nucleon and collective motions.
Generally, at low spins, the properties of the $A\approx 60-70$ nuclei are governed by single-particle excitations involving the $1f_{5/2}$, $2p_{3/2}$, and $2p_{1/2}$ orbitals.
At intermediate and high spins, the deformation-driving $1g_{9/2}$ orbitals come into play and induce the collective behavior.
Because of this complex interplay, many interesting phenomena have been observed in this mass region, such as the superdeformed bands \cite{Rudolph1998PRL,Ward2000PRC,Johansson2009PRC} and the magnetic rotations \cite{Torres2008PRC,Steppenbeck2012PRC}.

Even though studies on the even-even nuclei in the $A\approx 60-70$ mass region were widely performed \cite{ZhuS2006PRC,Ljungvall2010PRC,Hoteling2010PRC,Crawford2013PRL,Albers2016PRC,Sensharma2022PRC,Ayangeakaa2022PRC}, studies on the odd-mass nuclei were relatively scarce.
It is worthwhile to explore the spectroscopic properties of odd-mass nuclei,
since the existence of the unpaired nucleon would result in additional complexity and more interesting physics.
In fact, studies of these odd systems have provided valuable insights in the evolution of the single-particle states around the $Z=28$ closed shell \cite{Franchoo1998PRL,Stefanescu2008PRL} and the effects of collective motions \cite{Albers2013PRC,Ayangeakaa2015PRC}.

With three protons beyond the $Z=28$ closed shell, the odd-mass Ga isotopes present diverse spectral structures and has caught much attention \cite{Ivascu1974NPA,Riccato1974NPA,Stefanescu2009PRC,Diriken2010PRC}.
In particular, many experimental efforts have been devoted to measure the spectra of $^{69}$Ga since 1956 \cite{Fagg1956PR}.
These investigations reported negative-parity states generated by single-particle excitations involving $1f_{5/2}$, $2p_{3/2}$, $2p_{1/2}$ and $1f_{7/2}$ orbitals \cite{Temperley1965PR,Langhoff1968NPA,Moreh1973PRC,Paradellis1978PRC,Bakoyeorgos1982PRC}.
Very recently, through the $^{26}\text{Mg}(^{48}\text{Ca},p4n\gamma)$ multi-nucleon transfer reaction, the level scheme of $^{69}$Ga has been considerably extended.
In particular, three positive-parity bands S\uppercase\expandafter{\romannumeral1}, S\uppercase\expandafter{\romannumeral2}, and S\uppercase\expandafter{\romannumeral3} were observed in $^{69}$Ga for the first time \cite{Idoko2025}.

Theoretically, the spectroscopic properties of nuclei in the $A\approx 60-70$ mass region has been investigated with many theoretical approaches, e.g., the interacting boson-fermion model \cite{Timar1993NPA,Danko1999PRC}, the configuration-interaction shell model \cite{Poves2001NPA,Honma2004PRC}, and the tilted-axis-cranking covariant density functional theory (TAC-CDFT) \cite{ZhaoPW2011PLB,PengJ2018PRC,LuoDW2022PRC}.
The TAC-CDFT \cite{MengJ2013FP,ZhaoPW2011PRL} starts from a universal density functional and, thus, could provide a microscopic understanding on the nuclear single-nucleon and collective motions without introducing adjustable parameters. 
Since the TAC-CDFT takes Lorentz symmetry into account, it offers a consistent treatment of nuclear currents, which is quite important in the description of nuclear rotations \cite{Koepf1989NPA,Konig1993PRL,Afanasjev2000PRC,MengJ2013FP}.

In this work, the TAC-CDFT is applied to investigate the three newly-observed positive-parity bands S\uppercase\expandafter{\romannumeral1}, S\uppercase\expandafter{\romannumeral2}, and S\uppercase\expandafter{\romannumeral3} in $^{69}$Ga.
A microscopic interpretation for the collective structures in $^{69}$Ga will be provided.

\section{Theoretical Framework}

The detailed introduction of the CDFT can be found in Refs. \cite{Ring1996PPNP,Vretenar2005PR,MengJ2006PPNP,MengJ2016Relativistic}.
The detailed formalism of the TAC-CDFT can be found in Refs. \cite{PengJ2008PRC2,ZhaoPW2011PRL,ZhaoPW2011PLB,MengJ2013FP}. Here, a brief introduction is presented.

The starting point of the CDFT is a universal density functional \cite{Ring1996PPNP,Vretenar2005PR,MengJ2016Relativistic,MengJ2016Relativistic}.
To realize the TAC calculations based on the CDFT, the functional is transformed into a body-fixed frame rotating with a constant angular velocity $\boldsymbol{\omega}$ around an orientation which is not parallel to the principal axes.
The relativistic Kohn-Sham equation in the rotating body-fixed frame reads,
\begin{equation}
\left[\boldsymbol{\alpha}\cdot(-i\boldsymbol{\nabla}-\boldsymbol{V})+\beta(m+S)+V-\boldsymbol{\omega}\cdot{\boldsymbol{\hat{J}}}\right]\psi_k=\varepsilon_k\psi_k
\end{equation}
where ${\boldsymbol{\hat{J}}}$ is the nuclear total angular momentum. The scalar field $S$ and vector field $V^{\mu}$ are connected
in a self-consistent way to the nucleon densities and currents \cite{ZhaoPW2011PLB}.
The iterative solution of these equations yields single-particle energies, expectation values of three components $\langle J_i\rangle$ of the angular momentum, total binding energies, $B(E2)$ transition probabilities, etc. The magnitude of the angular velocity $\boldsymbol{\omega}$ is connected to the total angular momentum quantum number $I$ by the semiclassical relation $\langle \hat{\boldsymbol{J}}\rangle\cdot\langle \hat{\boldsymbol{J}}\rangle=I(I+1)$.

In the TAC-CDFT framework, the pairing correlations are often treated by the Bogoliubov method \cite{Vretenar2005PR,MengJ2016Relativistic} and the shell-model-like approach (SLAP). Compared to the Bogoliubov method, the SLAP treats the pairing correlations exactly by diagonalizing the many-body Hamiltonian in a many-particle configuration (MPC) space and has the advantage to keep the particle number conservation and avoid the pairing collapse, for details, see Refs. \cite{ZengJY1983NPA,MengJ2006FPC,WangYP2024PRL,XuFF2024PRL}.





Here, to investigate the collective structure in $^{69}$Ga, the single-particle Dirac equation is solved using a three-dimensional harmonic-oscillator basis in Cartesian coordinates with ten major shells, ensuring convergence for nuclei in the $A\approx 60-70$ mass region \cite{ZhaoPW2011PLB}. 
The PC-PK1 relativistic density functional \cite{ZhaoPW2010PRC} is adopted. 
Such a functional has demonstrated high predictive power in describing nuclear masses \cite{YangYL2021PRC,ZhangKY2022ADNDT} and shapes \cite{XuFF2024PRC,XuFF2024PLB,YangYL2023PRC}, magnetic and antimagnetic rotations \cite{ZhaoPW2011PRL,ZhaoPW2011PLB,MengJ2016PS,WangYK2017PRC,WangYK2018PRC}, chiral rotation \cite{ZhaoPW2017PLB,WangYK2019PRC,RenZX2020PRC,WangYP2023PLB}, and nuclear fission \cite{RengZX2022PRL,LiB2023PRC}, etc.

\section{Results and discussions}

The occupation of the $g_{9/2}$ orbitals plays an important role in describing the collective structures observed in $^{69}$Ga. 
To identify suitable configurations involving these $g_{9/2}$ orbitals for the positive-parity bands S\uppercase\expandafter{\romannumeral1}, S\uppercase\expandafter{\romannumeral2}, and S\uppercase\expandafter{\romannumeral3}, cranking calculations using the configuration constrained approach \cite{MengJ2013FP} are performed when the pairing correlations are not taken into account. 
By comparing the obtained results with the experimental data, the two most likely configurations are identified. 
The calculated energies and deformation parameters $\beta$ and $\gamma$ for these two configurations at $\hbar\omega = 0$ MeV are listed in Table \ref{tab1}.
The valence-nucleon configuration for band S\uppercase\expandafter{\romannumeral1} is determined to be $\nu(g_{9/2})^4(pf)^6\otimes \pi(g_{9/2})^1(pf)^2$.
The configuration $\nu(g_{9/2})^4(pf)^6\otimes\pi(g_{9/2})^3$ with positive and negative signatures is assigned to bands S\uppercase\expandafter{\romannumeral2} and S\uppercase\expandafter{\romannumeral3}, respectively.
The $\gamma$ deformations for these two configurations are zero, which corresponds to the prolate shape.
The corresponding $\beta$ deformations are large for both of the configurations, 
which are expected to be related to the valence $g_{9/2}$ proton and neutron orbitals. 
The particle nature of these configurations generates principal axis rotations, which are consistent with the $\Delta I=2$ character of the bands under consideration.

\begin{table}[htbp!]
  \renewcommand{\arraystretch}{1.5}
  \setlength{\tabcolsep}{2.5pt}
  \centering
  \caption{Energies, deformations $\beta$ and $\gamma$, and configurations for positive-parity bands S\uppercase\expandafter{\romannumeral1}-S\uppercase\expandafter{\romannumeral3} in $^{69}$Ga.}
  \begin{tabular}{ccccc}
  \toprule
  Band & $E$ (MeV) & $\beta$ & $\gamma$ & Configuration\\
  \midrule
  S\uppercase\expandafter{\romannumeral1} & -596.2 & 0.41 & $0.0^{\circ}$ & $\nu(g_{9/2})^4(pf)^6\otimes \pi(g_{9/2})^1(pf)^2$\\
  S\uppercase\expandafter{\romannumeral2}, S\uppercase\expandafter{\romannumeral3} & -595.6 & 0.57 & $0.0^{\circ}$ & $\nu(g_{9/2})^4(pf)^6\otimes\pi(g_{9/2})^3$\\
  \bottomrule
  \end{tabular}
  \label{tab1}
\end{table}

\begin{figure}[htbp!]
  \centering
  \includegraphics[width=0.95\linewidth]{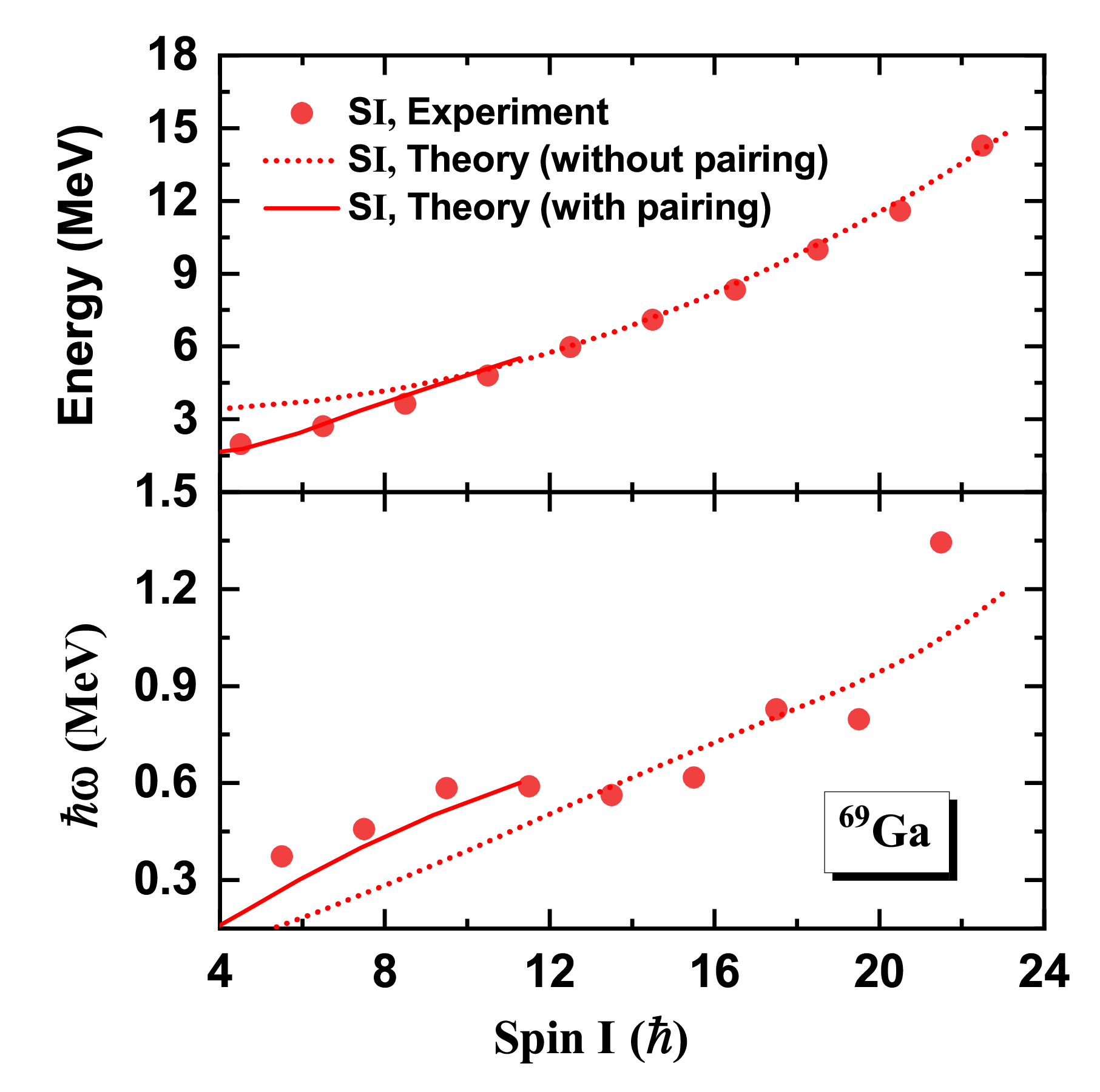}
  \caption{Calculated rotational energy (upper panel) and rotational frequency (lower panel) as functions of the angular momenta for the positive-parity band S\uppercase\expandafter{\romannumeral1} in $^{69}$Ga with and without pairing, in comparison with the data \cite{Idoko2025}.}
  \label{fig1}
\end{figure}

The calculated energy spectra and the angular momenta for the positive-parity band S\uppercase\expandafter{\romannumeral1} are shown in Fig. \ref{fig1}, in comparison with the experimental data \cite{Idoko2025}.
Without taking into account the pairing correlations, the theoretical results give a good description of the states with spin $I> 23/2\hbar$.
Nevertheless, the observed energies and the total angular momenta for the states with $I \leq 23/2\hbar$ are overestimated, which might be related to the lack of pairing correlations.
To better clarify this point, the SLAP is applied to treat the pairing correlations in the framework of TAC-CDFT.  
In the calculations, the monopole pairing force is adopted.
The effective neutron and proton pairing strengths are respectively 0.7 MeV and 0.75 MeV, determined by reproducing the experimental odd-even mass differences. 
The many-particle configuration (MPC) space is truncated at an excitation energy cutoff \cite{WuCS1989PRC} of $E_c=10$MeV.
A larger MPC space with a renormalized pairing strength gives essentially the same results, indicating the convergence of the MPC space.
After the inclusion of the pairing correlations, it can been seen that both the calculated energy spectra and the total angular momenta are in good agreement with the experimental data.


\begin{figure}[htbp!]
  \centering
  \includegraphics[width=0.95\linewidth]{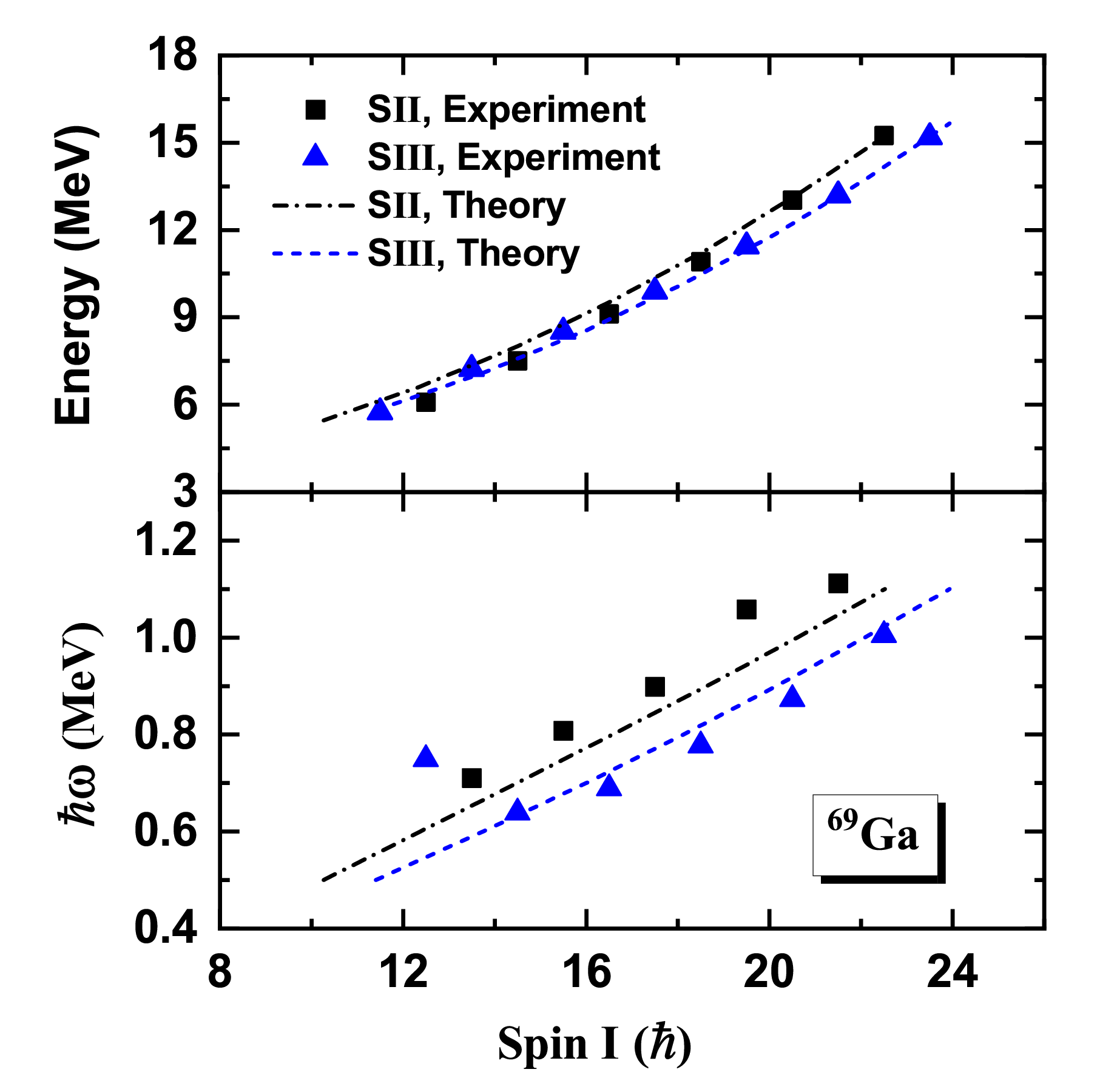}
  \caption{Calculated rotational energy (upper panel) and rotational frequency (lower panel) as functions of the angular momenta for the positive-parity bands S\uppercase\expandafter{\romannumeral2} and S\uppercase\expandafter{\romannumeral3} in $^{69}$Ga, in comparison with the data \cite{Idoko2025}.}
  \label{fig2}
\end{figure}

The calculated energy spectra and the angular momenta for the positive-parity bands S\uppercase\expandafter{\romannumeral2} and S\uppercase\expandafter{\romannumeral3} in $^{69}$Ga as well as their comparison with the data \cite{Idoko2025} are shown in Fig. \ref{fig2}.
The spins for the band heads of bands S\uppercase\expandafter{\romannumeral2} and S\uppercase\expandafter{\romannumeral3} are already larger than $10\hbar$, leading to the fact that the pairing correlations are negligible for these two bands.
The experimental energies and angular momenta for both bands S\uppercase\expandafter{\romannumeral2} and S\uppercase\expandafter{\romannumeral3} are properly reproduced by the calculations without the pairing correlations.
Since these two bands are based on the same configuration with opposite signatures, they are suggested as signature partner bands, in which the one with the negative signature is more energy favored.


\begin{figure}[htbp!]
  \centering
  \includegraphics[width=0.95\linewidth]{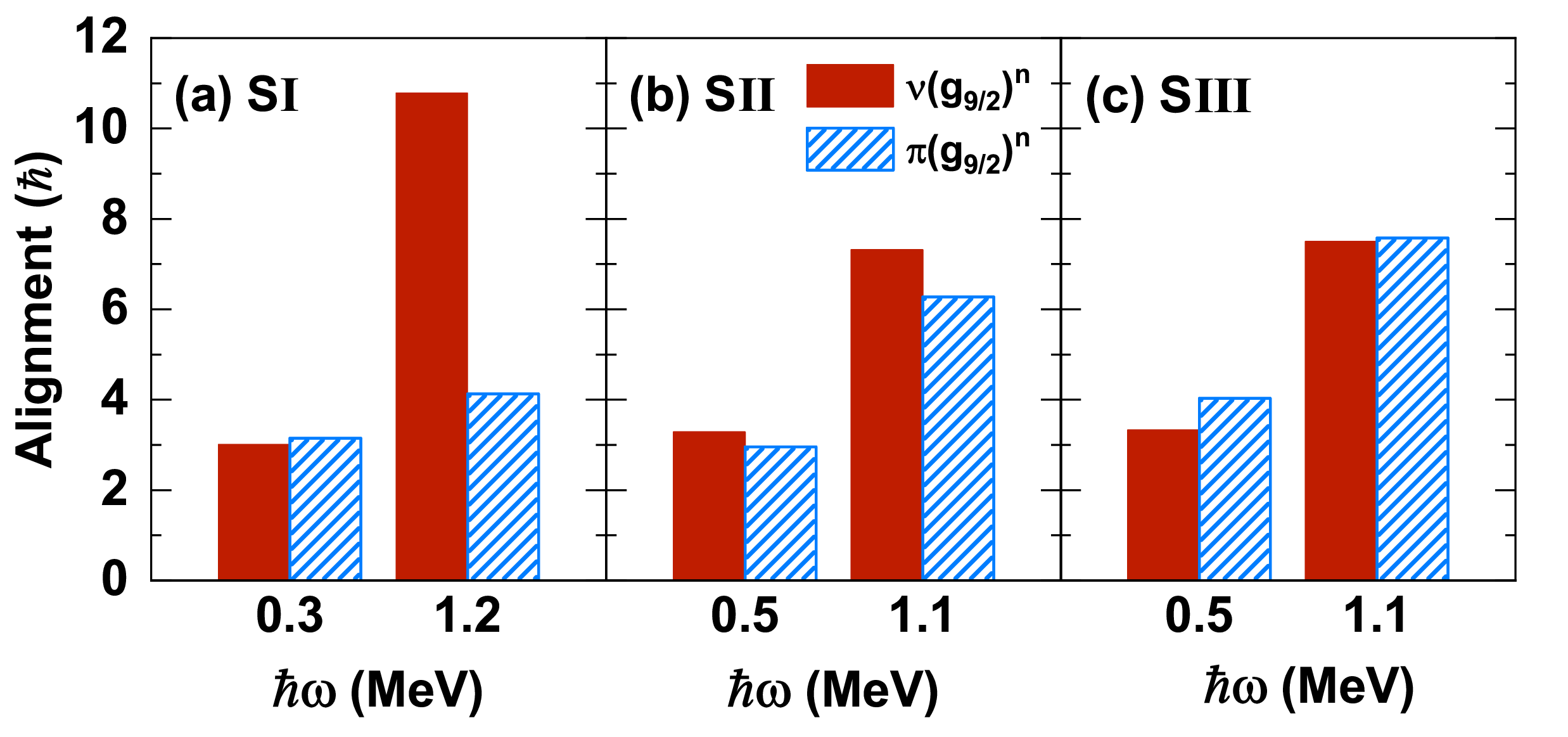}
  \caption{Angular momentum alignments of the proton and neutron particles in the $g_{9/2}$ shell for the bands S\uppercase\expandafter{\romannumeral1} (a), S\uppercase\expandafter{\romannumeral2} (b) and S\uppercase\expandafter{\romannumeral3} (c), calculated by the TAC-CDFT without the pairing correlations. The numbers below the abscissa denote the rotational frequency at which the plotted alignments have been obtained.}
  \label{fig5}
\end{figure}

As mentioned before, all the studied bands involves nucleons in $g_{9/2}$ orbits, and the reasonable description of these bands gives us confidence to give a microscopic interpretation of the role of the $g_{9/2}$ orbitals in the collective structures in $^{69}$Ga. 
In the present microscopic calculations, the nuclear total angular momenta are generated from all the individual nucleons inside of a self-consistent meanfield potential. 
For band S\uppercase\expandafter{\romannumeral1}, there are four $g_{9/2}$ valence neutrons and one $g_{9/2}$ valence proton.
For bands S\uppercase\expandafter{\romannumeral2} and S\uppercase\expandafter{\romannumeral3}, there are four $g_{9/2}$ valence neutrons and three $g_{9/2}$ valence protons.
The calculated angular momentum alignments of these $g_{9/2}$ nucleons calculated by the TAC-CDFT without the pairing correlations are shown in Fig. \ref{fig5}.
For band S\uppercase\expandafter{\romannumeral1} as shown in Fig. \ref{fig5} (a), the contribution from the two neutrons in the $g_{9/2}$ orbits increases with $\omega$, and make a principal contribution to the generation of the total angular momentum.
In contrast, the alignment of the unpaired $g_{9/2}$ proton is almost constant.
For band S\uppercase\expandafter{\romannumeral2} as shown in Fig. \ref{fig5} (b), the alignments of the valence $g_{9/2}$ neutrons and protons increase simultaneously with $\omega$  and their contributions to the total angular momentum are about the same order.
The behavior of band S\uppercase\expandafter{\romannumeral3} is similar with its signature partner band S\uppercase\expandafter{\romannumeral2} and is shown in Fig. \ref{fig5} (c).



\begin{figure*}[htbp!]
  \centering
  \includegraphics[width=0.95\linewidth]{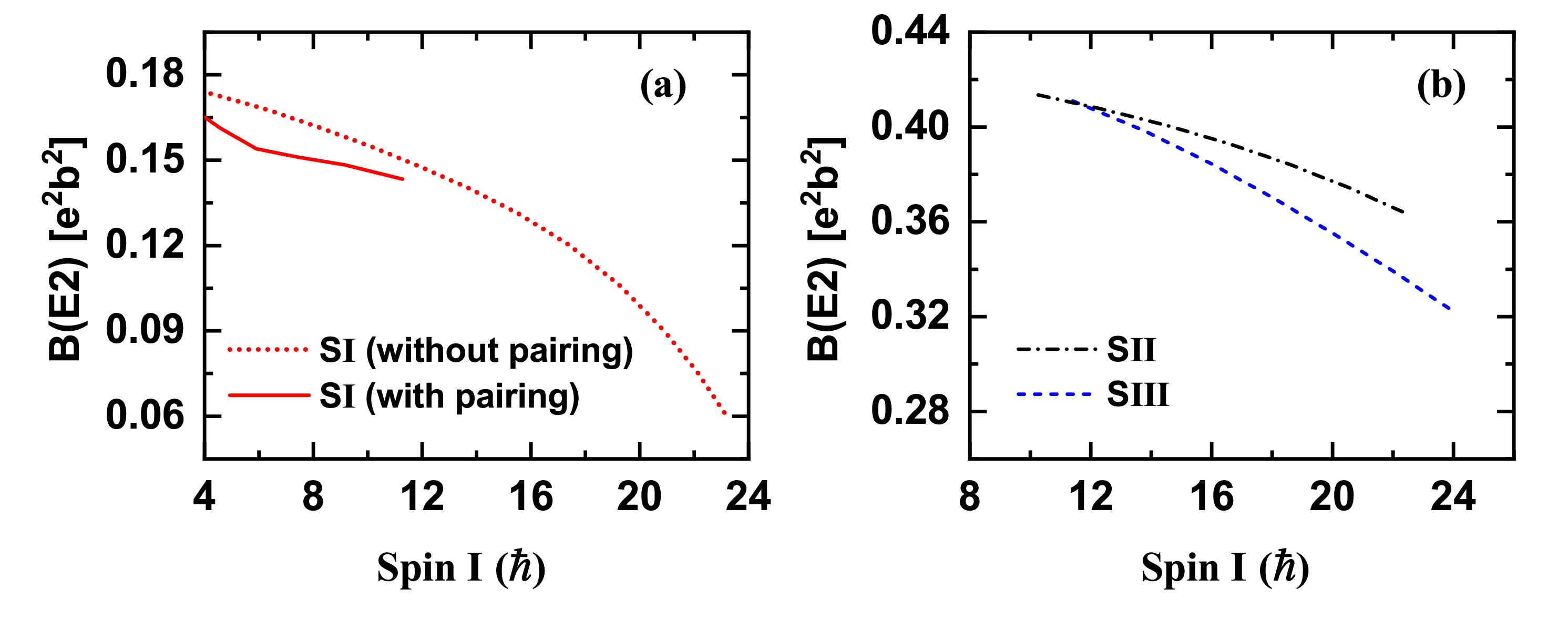}
  \caption{Calculated $B(E2)$ values as functions of the angular momenta for the positive-parity bands (a) S\uppercase\expandafter{\romannumeral1}, as well as (b) S\uppercase\expandafter{\romannumeral2} and S\uppercase\expandafter{\romannumeral3}.}
  \label{fig3}
\end{figure*}

Finally, the calculated B(E2) values for the positive-parity bands S\uppercase\expandafter{\romannumeral1}, S\uppercase\expandafter{\romannumeral2} and S\uppercase\expandafter{\romannumeral3} are shown as functions of the angular momenta in Fig. \ref{fig3}. 
The $B(E2)$ values exhibit a decreasing tendency for all these bands.
This is caused by the fact that their deformation parameters $\beta$ decrease with the total angular momenta.
For band S\uppercase\expandafter{\romannumeral1}, when taking into account the pairing correlations, the $B(E2)$ values for the states with spin $I \leq 23/2\hbar$ are slightly reduced.
For the predicted signature partner bands S\uppercase\expandafter{\romannumeral2} and S\uppercase\expandafter{\romannumeral3}, the calculated $B(E2)$ values are close at spin $I \approx 23/2\hbar$. 
At higher spins, however, the $B(E2)$ value of band S\uppercase\expandafter{\romannumeral2} becomes larger than those of band S\uppercase\expandafter{\romannumeral3}.
Additional experimental efforts including the measurement of lifetimes of these states are required to validate the predicted transition probabilities.

\section{summary}

In summary, the tilted-axis-cranking covariant density functional theory (TAC-CDFT) is applied to investigate the three newly-observed positive-parity bands S\uppercase\expandafter{\romannumeral1}, S\uppercase\expandafter{\romannumeral2}, and S\uppercase\expandafter{\romannumeral3} in $^{69}$Ga.
The energy spectra and angular momenta are calculated and compared with the experimental data.
For the yrast band S\uppercase\expandafter{\romannumeral1}, pairing correlations play a crucial role for the states with spin $I\leq 23/2\hbar$.
The bands S\uppercase\expandafter{\romannumeral2} and S\uppercase\expandafter{\romannumeral3} are suggested to be signature partner bands with positive and negative signatures, respectively.
By analyzing the angular momentum alignments microscopicly, it is revealed that the $g_{9/2}$ protons and neutrons play an important role in the description of the collective structure of $^{69}$Ga.
For band S\uppercase\expandafter{\romannumeral1}, the two neutrons in the $g_{9/2}$ orbitals make a principal contribution to the generation of the total angular momentum.
For bands S\uppercase\expandafter{\romannumeral2} and S\uppercase\expandafter{\romannumeral3}, the alignments of the valence $g_{9/2}$ neutrons and protons increase simultaneously with $\omega$ and their contributions to the total angular momentum are about the same order.
The transition probabilities $B(E2)$ for these bands are predicted, and await further experimental verification.

\begin{acknowledgments}

This work was partly supported by the National Natural Science Foundation of China (Grants No. 12435006, No. 12475117, No. 12141501, No. 11935003), the High-performance Computing Platform of Peking University, and the National Key Laboratory of Neutron Science and Technology (Grant No. NST202401016).

\end{acknowledgments}


\end{document}